# Large Nernst Power Factor over a Broad Temperature Range in Polycrystalline Weyl Semimetal NbP

Chenguang Fu,*‡[a]  Satya N. Guin,‡[a] Sarah J. Watzman,[a,b] Guowei Li,[a] Enke Liu,[a] Nitesh Kumar,[a] Vicky Süß,[a] Walter Schnelle,[a] Gudrun Auffermann,[a] Chandra Shekhar,[a] Yan Sun,[a] Johannes Gooth,[a] and Claudia Felser*[a]

[a]Max Planck Institute for Chemical Physics of Solids, Nöthnitzer Str. 40, 01187 Dresden, Germany.

[b]Department of Mechanical and Aerospace Engineering, The Ohio State University, Columbus, OH 43210, USA.

E-mail: Chenguang.Fu@cpfs.mpg.de, Claudia.Felser@cpfs.mpg.de

The discovery of topological materials has provided new opportunities to exploit advanced materials for heat-to-electricity energy conversion as they share many common characteristics with thermoelectric materials. In this work, we report the magneto-thermoelectric properties and Nernst effect of the topological Weyl semimetal NbP. We find that polycrystalline, bulk NbP shows a significantly larger Nernst thermopower than its conventional thermopower under magnetic field. As a result, a maximum Nernst power factor of ~ 35×10-4 Wm-1K-2 is achieved at 9 T and 136 K, which is 4 times higher than its conventional power factor and is also comparable to that of state-of-the-art thermoelectrics. Moreover, the Nernst power factor maintains relatively large value over a broad temperature range. These results highlight that the enhancement of thermoelectric performance can be achieved in topological semimetals based on the Nernst effect and transverse transport.

## Introduction

Thermoelectric (TE) materials, which can directly convert waste-heat into electricity and also work as solid-state refrigerators, provide a potential solution for energy conversion and have attracted vast research activities over the past century.[1-5] The basis of current TE materials research is the Seebeck effect, a phenomenon in which an applied temperature gradient generates a parallel voltage between the hot and cold sides of the conductive material (Fig. 1a).[6] Both electrons and holes in a conductive solid condense on the cold side; their opposite charges counterbalance each other's contribution to the induced TE voltage, leading to a reduced thermopower (or Seebeck coefficient) in two-carrier systems.  Thus, state-of-the-art TE materials are generally doped semiconductors with one dominant carrier, either electrons ($n$-type) or holes ($p$-type).[7] Strategies that suppress the bipolar effect to maintain a high thermopower, such as broadening the band gap or raising the density of majority carriers, have been widely employed to improve TE performance.[8-10] In contrast, semimetals, which naturally possess two types of carriers, are usually attributed as bad thermoelectric materials and have attracted less attention in TE research.

If a conductive solid is exposed to a magnetic field mutually perpendicular to the applied temperature gradient, charge carriers deflect due to the Lorentz force, generating a transverse electric field orthogonal to both the applied magnetic field and temperature gradient (Fig. 1b). This is the Nernst effect, a type of thermomagnetic (TM) effect, first discovered by Ettingshausen and Nernst in a study of the compensated semimetal Bi.[11] Distinct from the Seebeck effect, in the Nernst effect, the magnetic field forces electrons and holes to deflect in opposite directions. Thus, both types of carriers contribute to the transverse TE voltage. Here, the ratio of the generated transverse voltage to the perpendicularly-applied magnetic field is called the Nernst thermopower. Therefore, semimetals, especially compensated ones, could be promising candidates for transverse energy conversion via the Nernst-Ettingshausen effect.

Past decades have witnessed tremendous achievements in TE materials based on the Seebeck effect.[1-5] However, TM materials based on the Nernst effect have attracted much less attention.[12] This is, in part, due to the necessity of an externally applied magnetic field. Today, permanent magnets allow us to easily obtain a moderate magnetic field, giving potential for TM materials in energy conversion applications.

In general, the combination of a small Fermi surface and long carrier mean-free-path in a compensated semimetal can generate a large Nernst signal, such as in Bismuth and graphite.[12,13] Recently, the theoretical prediction and experimental realization of topological Weyl and Dirac semimetals have generated enormous research interest.[14-17] These newly discovered topological semimetals exhibit crossing of linear bands, ultrahigh carrier mobility, and giant magnetoresistance.[17-19] Thus, they provide a functional material database for studying the Nernst effect and their TM properties. However, there are still very few experiments focusing on the Nernst effect of topological semimetals. Recently, Jia *et al.* and Liang *et al.* reported the Nernst effect of the Dirac semimetal $Cd_3As_2$.[20, 21] Watzman *et al.* found a large Nernst thermopower in the single-crystal Weyl semimetal NbP.[22] The anomalous Nernst effect was reported in the chiral antiferromagnetic $Mn_3Sn$ by Ikhlas *et al.* and Li *et al.*[23, 24] All aforementioned work focused on single crystals, which is beneficial for recognizing physical origin but is not the best avenue for large-scale materials performance exploration and applications. Experimental work with a focus on studying the potential of polycrystalline topological semimetals in energy conversion and solid-state cooling applications utilizing the Nernst effect are rare. Compared to single crystals, polycrystalline bulk materials have many advantages, such as being easy to synthesize, exhibiting a low thermal conductivity, and having an easily tunable Fermi level which facilitates rapid screening of functional materials. These facts motivate us to study the Nernst effect and TM properties of polycrystalline topological semimetals.

In this work, we show that polycrystalline Weyl semimetals are potential candidates for heat-to-electricity energy conversion and solid-state refrigeration. As a case study, we choose the Weyl semimetal NbP which displays large magnetoresistance, ultrahigh mobility and the mixed axial-gravitational anomaly.[17, 25] Here we investigate the Nernst effect and magneto-thermoelectric properties of this system. It is found that polycrystalline bulk NbP exhibits a larger Nernst thermopower $\alpha_{xy}$ than its conventional Seebeck thermopower $\alpha_{xx}$, resulting in a maximum Nernst power factor of ~ $35\times10^{-4}$ Wm$^{-1}$K$^{-2}$ at 9 T and 136 K, defined as $\alpha_{xy}^2\sigma_{yy}$, where $\sigma_{yy}$ is the electrical conductivity. This value is remarkably high, 4 times larger than its conventional power factor $\alpha_{xx}^2\sigma_{xx}$, and is comparable to that of state-of-the-art thermoelectrics. Thus, NbP is a potential candidate for energy conversion or solid-state refrigeration in a transverse geometry.

**Experimental Section**

Polycrystalline powder of NbP was first synthesized by a direct reaction of niobium (Chempur, 99.9%) and red phosphorus (Heraeus 99.999%) kept in an evacuated fused silica tube for 48 h at 800 °C. The obtained powders were placed in the graphite dies with an inner diameter of 10 mm and compacted by spark plasma sintering (SPS) instrument (SPS-515 ET, Fuji, Japan). Due to the lack of sintering parameters available in the literature for SPS of NbP, we tried different sintering temperatures under a constant uniaxial pressure of 80 MPa to compact the powders. The sintering temperature-dependent relative density of bulk NbP samples is shown in Table S1 (ESI†). For the current study, we chose the sample with the highest relative density (91%), which was fabricated in the following process: the graphite die with NbP powders was first quickly heated to 1100 °C and then slowly heated to 1150 °C, where it stayed for 5 min. A uniaxial pressure of 80 MPa and high vacuum were maintained during the whole sintering process. A small piece of NbP was cut from the bulk sample and grinded for power X-ray diffraction (XRD) measurement, which was performed at room temperature using a Huber Image Plate Guinier Camera G670 operated with Cu K$_{\alpha1}$ radiation (λ = 1.54056 Å). No obvious impurity phase was observed, as shown in Fig. S1 (ESI†). The microstructure and composition of the NbP sample were examined by the scanning electron microscopy with an energy dispersive X-ray spectroscopy analyzer, as shown in Fig. S2 and Table S2 (ESI†). The actual composition is in agreement with the nominal one when the instrument error is considered. For transport measurement, the bulk sample was cut into pieces with dimensions of 8.5 × 2 × 2 mm$^3$ (Fig. S3). The conventional thermopower above room temperature was measured by using an ULVAC ZEM-3 system. The temperature dependent thermal conductivity and Seebeck thermopower under magnetic field up to 9 T were simultaneously measured adiabatically by the one-heater and two-thermometer configuration using the thermal transport option (TTO) of the PPMS (Quantum Design) in which the sample was placed in an orientation where the magnetic field was perpendicular to the heat flow. The estimated error for Seebeck thermopower is ± 5%. For thermal conductivity, as the geometry could introduce additional uncertainty, a typical error is ± 10%. Both longitudinal and Hall resistivities under magnetic field were measured by a standard four-probe method using the PPMS. The accuracy of resistivity measurement is ± 3%. To correct for contact misalignment, the measured raw data were field-symmetrized and antisymmetrized, respectively. Nernst thermoelectric measurements were also conducted by the one-heater and two-thermometer configuration using the PPMS. The instrument was controlled by software programmed using LabVIEW. The measurement was done in the temperature range 18 - 312 K, and magnetic fields were swept in both directions to a maximum magnitude of 9 T. To generate a temperature gradient, a resistive heater was connected to a gold-plated copper plate at one end of the sample. The thermal gradient, $\Delta T$ was applied along the longest direction of the sample. For the heat sink, a gold-plated copper plate was attached to the puck clamp. To measure the temperature gradient, two gold-plated copper leads were attached directly to the sample using silver epoxy along the thermal gradient direction. The temperature difference was typically near 1 % - 3 % of the sample temperature. The transverse voltage was measured by attaching two copper wires, orthogonal to the temperature gradient direction of the sample, using silver epoxy. The Nernst signal was estimated as $\alpha_{xy} = L_x V_y / L_y \Delta T_x$, where, $V_y$ is transverse electric voltage, $L_x$ is the distance between the two temperature leads, $L_y$ is the distance between the two voltage wires, and $\Delta T_x$ is the measured temperature difference. The typical error on the field dependent Nernst thermopower is ± 10%.

**Results and discussion**

NbP has a non-centrosymmetric crystal structure in a tetragonal lattice with space group $I4_1md$ (No. 109) and is isostructural to TaAs, TaP and NbAs.[17] The crystal structure is built up of NbP$_6$ and PNb$_6$ trigonal prisms as shown in Fig. 1c. Both Nb and P atoms occupy the 4a Wyckoff position with atomic coordinate of (0, 0, 0) and (0, 0, 0.416), respectively. The band structure of NbP near its Fermi level together with the champion TE material PbTe and the elemental semimetal Bi is schematically shown in Fig. 1c. PbTe is a semiconductor with small direct band gap. By doping with either dopants or acceptors, excellent TE performance is exhibited in this system.[26, 27] Elemental Bi is a typical, compensated semimetal with coexistence of quadratic electron and hole pockets, which are favorable to its high Nernst thermopower.[13] NbP has unique band characteristics by combining quadratic electron and hole pockets and additional electron pockets from linear Weyl bands. Linear or nearly linear energy dispersion is usually an origin of high mobility, as has been observed in the recently-discovered topological semimetals.[17, 18] These features make NbP a candidate for a large Nernst effect and good TM properties.

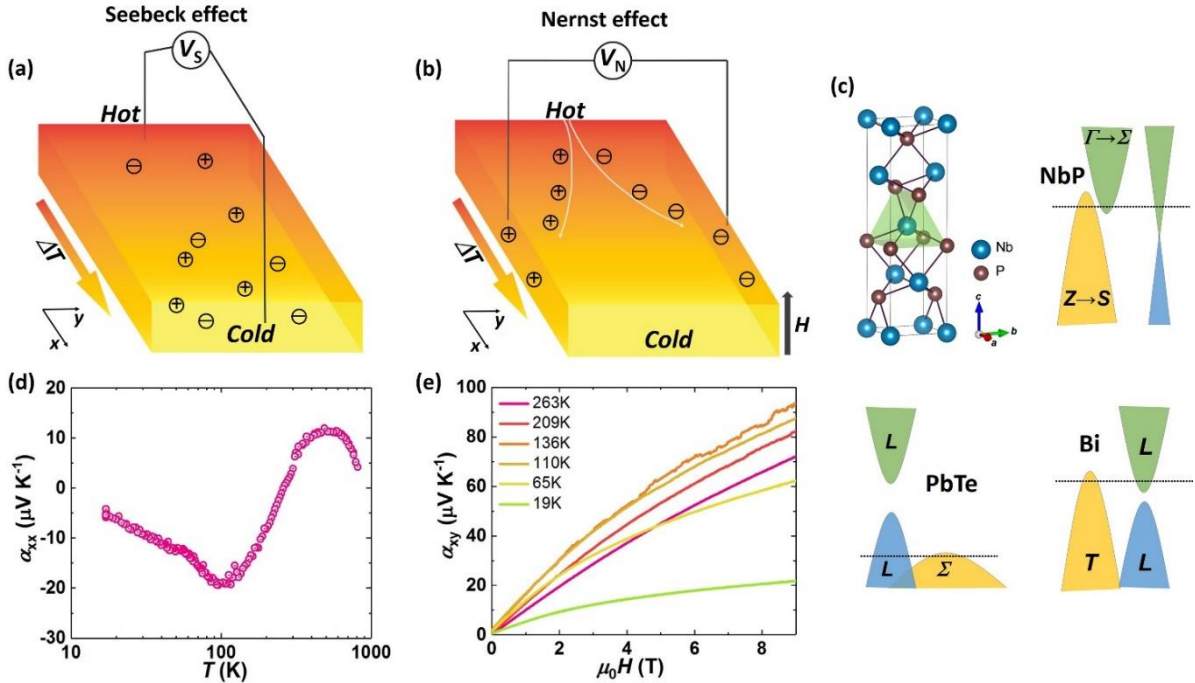

**Fig. 1** Schematic diagrams of Seebeck effect (a) and Nernst effect (b). $V_S$ and $V_N$ donate the longitudinal and transverse thermoelectric voltages, respectively. (c) Crystal structure of NbP (up left) and schematic illustrations of band structures for Weyl semimetal NbP (up right), semiconductor PbTe (down left) and elemental semimetal Bi (down right). The dash lines refer to the Fermi levels. Temperature dependent thermopower $\alpha_{xx}$ at 0 T (d) and Nernst thermopower $\alpha_{xy}$ versus magnetic field at different temperatures (e) for polycrystalline NbP.

Polycrystalline bulk NbP was successfully fabricated with the spark plasma sintering technique by compacting the microcrystalline powder, synthesized from a direct reaction of niobium and red phosphorus. Detailed information on synthesis and microstructure characterization can be found in the experimental section. The longitudinal thermopower (Seebeck coefficient) $\alpha_{xx}$ of polycrystalline NbP was measured at different temperatures, as shown in Fig. 1d. $\alpha_{xx}$ is positive at 0 T and above room temperature, signifying that holes dominate electrical transport here. Upon cooling, $\alpha_{xx}$ exhibits a sign reversal and reaches a minimum value of -20 μV K$^{-1}$ near 100 K. In a previous work, Watzman *et al.* reported a minimum $\alpha_{xx}$ of about -10 μV K$^{-1}$ in single crystal NbP.[22] The deviation of $\alpha_{xx}$ between single-crystalline and polycrystalline NbP might originate from the difference of the direction of measurement and slight change of the position of Fermi level induced by slightly different Nb/P stoichiometry. Overall, the obtained $\alpha_{xx}$ in both polycrystalline and single crystal NbP is about one order of magnitude lower than that of good TE materials, indicating the bad TE properties of NbP. In contrast, the transverse Nernst thermopower $\alpha_{xy}$ of polycrystalline NbP is superior to its thermopower. As displayed in Fig. 1e, with an applied magnetic field, $\alpha_{xy}$ of NbP increases rapidly and reaches an unsaturated value of about 90 μVK$^{-1}$ at 9 T and 136 K, which is comparable to the conventional thermopower of TE materials.

In Fig. 2, we display power factor (PF) and Nernst power factor (PF$_{Nernst}$) for polycrystalline NbP. As seen in Fig. 2a, at 0 T, the PF first increases and reaches a peak value of ~8.5 × 10$^{-4}$ Wm$^{-1}$K$^{-2}$ near 100 K then quickly falls down with increasing temperature. After applying a magnetic field, the peak PF moves to a higher temperature, but the peak value remains almost unchanged. In contrast, the PF$_{Nernst}$ (Fig. 2b) increases with increasing temperature and reaches a maximum near 150 K. Near 150 K, by further increasing the magnetic field, PF$_{Nernst}$ increases rapidly and achieves a peak value of ~35 × 10$^{-4}$ Wm$^{-1}$K$^{-2}$ at 9 T, which is about 4 times higher than the corresponding PF and is comparable to the PF of state-of-the-art thermoelectrics, such as Bi$_2$Te$_3$[28, 29], PbTe[26, 30], CoSb$_3$,[31, 32] and half-Heusler compounds[33, 34]. Generally, the heat source in real system is diffusive in nature and hence for an effective material it is more important to have high PF over a broad temperature range rather than a peak PF. The PF of polycrystalline NbP shows a clear peak value while the PF$_{Nernst}$ maintains relatively large value > 20 × 10$^{-4}$ Wm$^{-1}$K$^{-2}$ at 9 T over a broad temperature range from 100 K to 300 K. We calculate the average PF and PF$_{Nernst}$ of polycrystalline NbP, calculated using an integration way: $\left(\frac{1}{T_H-T_L}\right) \int_{T_L}^{T_H} PF(T)dT$, where $T_H$ and $T_L$ are the highest and lowest temperatures in this study, respectively (Fig. 2c). The average PF shows only a slight increase while the average PF$_{Nernst}$ displays dramatic enhancement with an applied magnetic field over a broad temperature range. The maximum average PF$_{Nernst}$ is about 25 × 10$^{-4}$ Wm$^{-1}$K$^{-2}$ at 9 T, 6 times higher than the corresponding average PF. This result highlights the potential of Weyl semimetals for TM applications at low temperature.

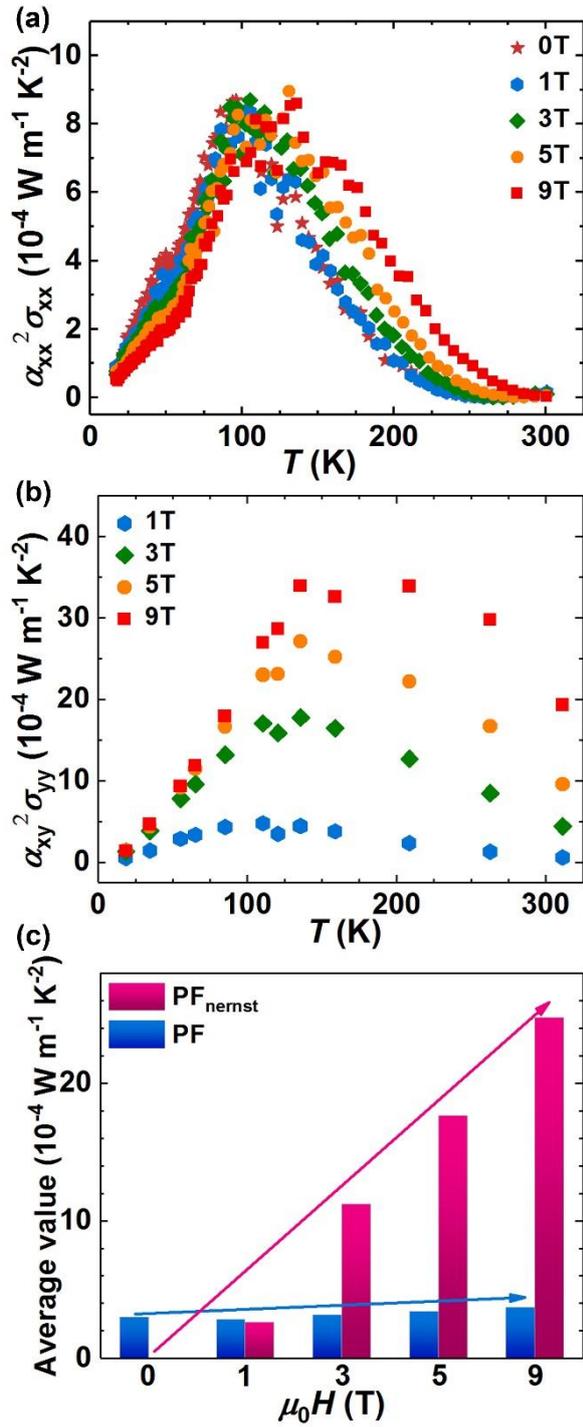

**Fig. 2** The power factor PF $\alpha_{xx}^2\sigma_{xx}$ (a), Nernst power factor PF$_{nernst}$ $\alpha_{xy}^2\sigma_{yy}$ (b) and average values of PF and PF$_{nernst}$ over the investigated temperature range (c) for polycrystalline NbP under different magnetic

To fully explore and compare the TM and TE properties of polycrystalline NbP, the thermopower, Nernst thermopower, and longitudinal and Hall resistivities are further investigated at different temperatures and magnetic fields, which are displayed in Fig. 3. $\alpha_{xx}$ decreases with applying a magnetic field and reaches a peak value of -47 μV K$^{-1}$ near 110 K and 9 T. A similar declining trend of field-dependent thermopower was reported in a recent study of single crystalline NbP.[35] Temperature-dependent Nernst thermopower is displayed in Fig. 3b, which is an odd function of magnetic field. The significant change of $\alpha_{xy}$ with magnetic field, which is maintained over a broad temperature range results in the large average PF$_{Nernst}$, as mentioned previously. The Nernst coefficient, defined as $N_{xy} = d\alpha_{xy}/d(\mu_0 H)$, is shown in Fig. S4 (ESI†). At low magnetic field (0 T < $\mu_0 H$ < 3 T), $N_{xy}$ is about twice as large as the value at higher field (3 T < $\mu_0 H$ < 9 T). Both at low and high fields, $N_{xy}$ peaks near 136 K. Furthermore, it is worth to discuss about that the obtained absolute $\alpha_{xx}$ and $\alpha_{xy}$ at high fields in the polycrystalline sample are smaller than those reported in the single crystals.[22,35] As per our understanding the large difference might be due to three factors: First, the polycrystalline nature of our sample makes the direction of measurement different from that of the single crystals. The grain boundaries in the polycrystalline sample might scatter the carriers and affect the transport properties. Second, as NbP has very small Fermi surface,[17] slight variation of stoichiometry in different samples could lead to shift of Fermi level and affect the transport properties. Third, the single crystals typically have much smaller size, which could lead large uncertainty of the measurement.

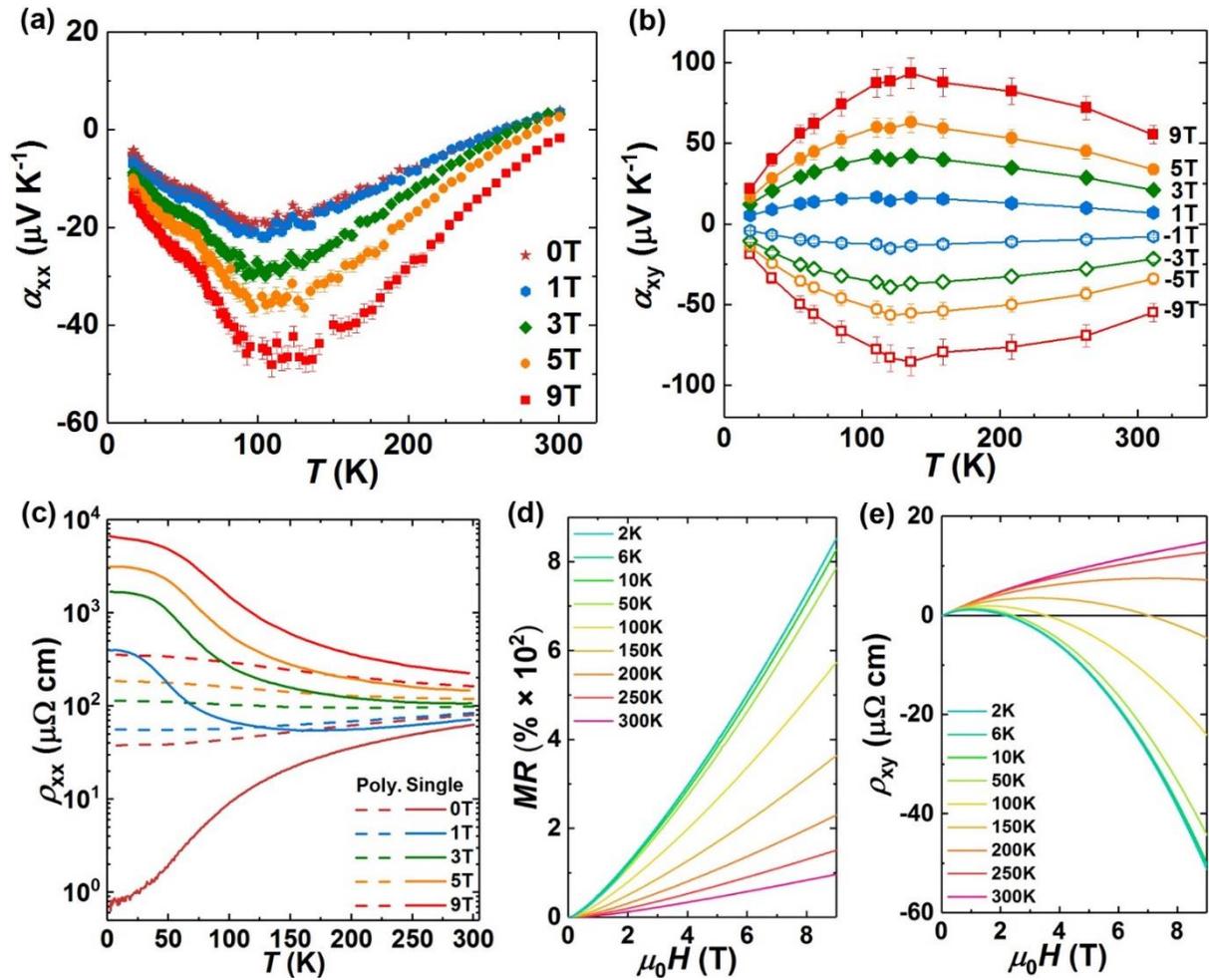

**Fig. 3** Electrical transport properties for polycrystalline NbP. Temperature dependence of thermopower $\alpha_{xx}$ (a) and Nernst thermopower $\alpha_{xy}$ (b) under different magnetic fields. (c) Longitudinal resistivity $\rho_{xx}$ under different magnetic fields comparing between polycrystalline and single-crystalline NbP samples. (d) Transverse magnetoresistance measured at different temperatures in magnetic fields up to 9 T. (e) Hall resistivity $\rho_{xy}$ at different temperatures.

Fig. 3c shows the temperature dependence of longitudinal resistivity $\rho_{xx}$ for polycrystalline NbP compared to that of single-crystalline NbP in magnetic fields up to 9 T. $\rho_{xx}$ of single-crystalline NbP is taken from literature for comparison[17]. At 0 T, $\rho_{xx}$ of polycrystalline NbP is much larger than that of single crystalline NbP. Especially at low temperature, the former is almost two orders of magnitude higher, indicating the strong grain boundary scattering and thus lower carrier mobility in polycrystalline NbP. However, on application of a magnetic field larger than 3 T, $\rho_{xx}$ of polycrystalline NbP is even lower than that of single crystalline NbP, which is favorable to a larger PF and $PF_{Nernst}$, as they are inversely proportional to $\rho_{xx}$. Furthermore, the transverse magneto-resistance (MR) of polycrystalline NbP is calculated using the formula $[\rho(\mu_0H) - \rho(0)] / \rho(0)$, as shown in Fig. 3d. At low temperatures, a relatively large MR = $8.5 \times 10^2$ % is found at 2 K in a field of 9 T. Moreover, at 300 K, a large MR of ~ $10^2$ % is still observed at 9 T, indicating the electrical transport of NbP can be significantly modulated by magnetic field at room temperature. The Hall resistivity, $\rho_{xy}$ of polycrystalline NbP is shown in Fig. 3e, which exhibits a nonlinear behavior at low fields, indicating that two types of carriers dominate the electrical transport properties. This explains the relatively low $\alpha_{xx}$ and typical semimetallic behavior of NbP. For a rough estimation of the Hall coefficient, carrier density and mobility, we use the slope of $\rho_{xy}$ at high fields (7 – 9T) based on the single-carrier Drude model, as Shekhar *et al.* did for single crystal NbP.[17] At 300 K and 2 K, the carrier density is calculated to be $5.5 \times 10^{20}$ cm$^{-3}$ and $5.4 \times 10^{19}$ cm$^{-3}$, respectively, while the corresponding carrier mobility is 142 cm$^2$V$^{-1}$s$^{-1}$ and 3085 cm$^2$V$^{-1}$s$^{-1}$, respectively. Compared to the carrier mobility of single-crystalline NbP, the polycrystalline sample is about 3 orders of magnitude lower, which results in the lower MR, as clearly observed in Fig. 3c. In contrast, at 300 K, the carrier mobility of polycrystalline sample is only about 45% lower, which explains why the MR is still smaller but approaching to that of single-crystalline NbP.

Thermal conductivity $\kappa_{xx}$ of polycrystalline NbP is shown in Fig. 4a, $\kappa_{xx}$ increases rapidly with increasing temperature and reaches a peak value of ~ 65 Wm$^{-1}$K$^{-1}$ near 85 K then decreases slowly to ~ 39 Wm$^{-1}$K$^{-1}$ at room temperature. It is worth noting that $\kappa$ of polycrystalline NbP is much lower than that of single crystalline NbP[22] due to enhanced grain boundary scattering. Furthermore, the magneto-thermal conductivity $\kappa_{xx}(\mu_0H)$ is measured up to 9 T, as shown in Fig. 4b, which displays an obvious declining trend with applied magnetic field. The reduction of $\kappa_{xx}(\mu_0H)$ is a result of the enhanced magneto-resistivity which leads to the reduction of electronic thermal conductivity. According to the magneto-thermal resistance (MTR) method [36, 37], $\kappa_{xx}(\mu_0H)$ can be rewritten in the form: $\kappa_{xx}(\mu_0H) = \kappa_e(\mu_0H) + \kappa_{ph}$, where $\kappa_e(\mu_0H)$ is the electronic thermal conductivity and defined as $LT\sigma_{xx}(\mu_0H)$ according to the Weidemann-Franz law, $L$ is defined as Lorenz number, $\kappa_{ph}$ is the phonon thermal conductivity and usually shows negligible change under magnetic field. As long as both $\kappa_{xx}(\mu_0 H)$ and $\sigma_{xx}(\mu_0 H)$ have the same functional form with respect to the magnetic field, they will have a linear relationship.[37] Thus, by plotting $\kappa_{xx}(\mu_0 H)$ against $\sigma_{xx}(\mu_0 H)$, one can extract the Lorenz number from the slope of the above formula; the intercept gives $\kappa_{ph}$. Fig. 4c shows a typical plot at 85 K; $\kappa_{xx}(\mu_0 H)$ and $\sigma_{xx}(\mu_0 H)$ follow a linear relationship, indicating the effectiveness of MTR method in analyzing the current data of polycrystalline NbP. Furthermore, by applying the MTR method to the data at other temperatures, we can extract the temperature-dependent $\kappa_{ph}$ and $\kappa_e(0)$, as shown in Fig. 4b. The extracted $L$ for all the investigated temperatures is in the range of $1.5 \sim 2.8 \times 10^{-8}$ W$\Omega$K$^{-2}$, which is near the metallic limit of $2.44 \times 10^{-8}$ W$\Omega$K$^{-2}$, verifying this method even though the experimental data has some uncertainty.

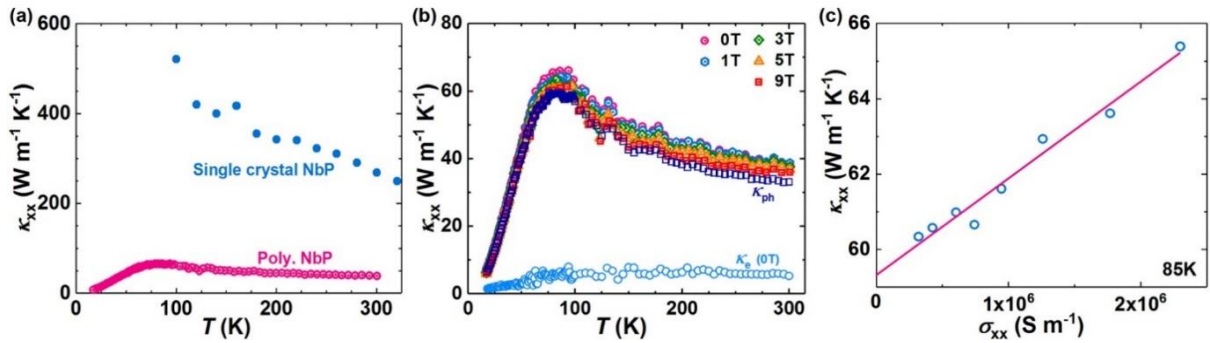

**Fig. 4** (a) Comparison of thermal conductivity between polycrystalline and single-crystalline NbP. The single-crystalline data is taken from the reference.[22] (b) Temperature dependence of thermal conductivity for polycrystalline NbP at different magnetic fields and the extracted phonon thermal conductivity $\kappa_{ph}$ and electronic thermal conductivity $\kappa_e$ by using the magneto-thermal resistance method. (c) Thermal conductivity is plotted against electrical conductivity. for polycrystalline NbP at 85 K. The red line shows a linear fitting of the experimental data.

Fig. 4b indicates that $\kappa_{ph}$ dominates the thermal conductivity in polycrystalline NbP. For example, at 300 K, $\kappa_{ph}$ is about 33 Wm$^{-1}$K$^{-1}$, contributing ~85 % to the total $\kappa$. $\kappa_{ph}$ of polycrystalline NbP is about one order of magnitude higher than that of good thermoelectric materials. With the measured electrical and thermal transport data, we also calculated the TE figure of merit $zT$ and Nernst $zT_{Nernst}$, which are defined as $\alpha_{xx}^2 \sigma_{xx} T / \kappa_{xx}$ and $\alpha_{xy}^2 \sigma_{yy} T / \kappa_{xx}$, respectively. As shown in Fig. S5 (ESI†), a peak $zT_{Nernst}$ of ~0.021 is achieved at 9 T and ~263 K, which is 8 times higher than its conventional peak $zT$. Although, the peak $zT_{Nernst}$ of polycrystalline NbP is still lower than the $zT$ of good TE materials, resulting from the intrinsically high $\kappa_{ph}$, as discussed above. However, the high PF$_{Nernst}$ and phonon-dominated thermal conduction are promising signatures for the optimization of the $zT_{Nernst}$ further. Fortunately, strategies aiming at the suppression of $\kappa_{ph}$ have been well established in the past decades, such as alloying, nanostructuring, or hierarchical phonon scattering.[30] Thus, by further suppressing the phonon thermal conductivity, one can expect a higher $zT_{Nernst}$ in polycrystalline NbP. Apart from NbP, there are many topological semimetals already discovered, and even many more have been predicted.[38] Very recently, Vergniory *et al.* checked 26,938 different stoichiometric materials from ICSD database, and they found about 10% of them are topological semimetals.[39] This finding provides great possibilities to seek potential candidates for energy conversion and solid-state refrigeration based on the Nernst-Ettingshausen effect. Among them, those with the combination of intrinsically low thermal conductivity and linear topological bands could be the most promising candidates.

## Conclusions

In summary, the polycrystalline bulk Weyl semimetal NbP has been successfully synthesized for the first time by a combination of direct reaction and spark plasma sintering process. The TE and TM properties under magnetic field are investigated in detail. We establish that the Nernst thermopower of polycrystalline NbP shows a larger value than its conventional thermopower. As a result, a maximum Nernst power factor of ~ $35 \times 10^{-4}$ Wm$^{-1}$K$^{-2}$ is achieved at 9 T, 136 K, which is 4 times higher than its conventional power factor and comparable to state-of-the-art thermoelectrics. More importantly, a remarkably high average Nernst power factor of ~ $25 \times 10^{-4}$ Wm$^{-1}$K$^{-2}$ is obtained below room temperature. The $zT_{Nernst}$ of the system can be further improved by phonon transport engineering. Additionally, seeking new topological semimetals with intrinsically low thermal conductivity could be an important direction to realize better TM performance. Our findings establish Weyl semimetal NbP as a model system for TM application and highlight the potential of topological semimetals for energy conversion and solid-state refrigeration.

## Conflicts of interest

The authors declare no competing financial interests.

## Acknowledgements

This work was financially supported by the ERC Advanced Grant No. (742068) "TOP-MAT" and funded by the Deutsche Forschungsgemeinschaft (DFG, German Research Foundation) - Projektnummer (392228380). C.G. Fu and S.N.G. acknowledge the financial support from Alexander von Humboldt Foundation. The authors thank Igor Veremchuk for assistance in using the SPS instrument.